\documentclass[12pt]{article}

\usepackage{graphicx}

\newcommand{\beq}{\begin{equation}}
\newcommand{\eeq}{\end{equation}}
\newcommand{\bea}{\begin{eqnarray}}
\newcommand{\eea}{\end{eqnarray}}
\newcommand{\noi}{\noindent}

\begin{document}

\centerline{\Huge Classicalization via Path Integral}

\vspace{1cm}

\centerline{Borut Bajc$^{a,b,}$\footnote{borut.bajc@ijs.si}, Arshad Momen$^{c,}$\footnote{amomen@univdhaka.edu} and Goran Senjanovi\' c$^{d,}$\footnote{goran@ictp.it}}

\vspace{0.5cm}
\centerline{$^{a}$ {\it\small J.\ Stefan Institute, Ljubljana, Slovenia}}
\centerline{$^{b}$ {\it\small Dept. of Physics, University of Ljubljana, Slovenia}}
\centerline{$^{c}$ {\it\small Dept. of Physics, University of Dhaka, Bangladesh}}
\centerline{$^{d}$ {\it\small ICTP, Trieste, Italy}}

\vspace{5mm}
   
\centerline{\large\sc Abstract}
\begin{quote}
\small

Recently, it was suggested that a large class of non-renormalizable
theories may need no UV completion. By analogy with gravity where classical 
black holes are expected to be created in high-energy scatterings, it is conjectured 
that similar classical solutions, so-called classicalons, should occur.  In this way the 
theory protects itself against non-unitarity, for instead of probing small distances at 
high energies one enters a classical regime. 
An effective theory of Goldstone bosons provides and example 
in which the size of classicalons grows with 
energy, and the high energy scattering is cut-off by small momenta, 
inversely proportional to the classicalon size. 
In this note we offer an alternative, path integral 
discussion of this important result.

\end{quote}

\vspace{7mm}

\section{Introduction}

With the advent of QED and then the Standard Model, renormalizability became 
the criterion for physically acceptable theories. Encouraged by the 
great successes 
of the work of Glashow, Weinberg and Salam, 
who followed the road of the UV completion (nowadays 
known as a Wilsonian approach) of the effective Fermi theory of weak interactions, 
this emerged as a canonical way of making sense out of non-renormalizable effective 
theories. 

The only trouble is that gravity still resists. The name of the game for now decades 
was and still is the search for its proper UV completion, the most popular one being  
strings. Recently a striking proposal~\cite{Dvali:2010bf} was put forward, according 
to which no UV completion whatsoever may be needed in the case of gravity. The 
essential point is both simple and deep. Normally one imagines that by going to very 
high energies $E$ one will probe small distances of the order $1/E$, and this is the 
crux of the problem in the case of non-renormalizable theories. The 
authors of~\cite{Dvali:2010bf} argue that by going to high energies, one will, as is 
well known, produce large classical black holes, with size

\beq
r_*=2G_NE
\eeq

\noi
and therefore one simply never probes small distances. For $E>>M_{Pl}$, 
$r_*>>1/M_{Pl}$. In other words, it is perfectly conceivable that no UV completion 
may be needed for $E>>M_{Pl}$, but that the theory simply enters the classical 
regime. The issue of how to compute the physical processes for $E\approx M_{Pl}$ 
remains an important unanswered question. Still it is remarkable if it were to 
turn true that gravity could have an in-built mechanism for self-protection 
against the problem of small distances.

Encouraged by this appealing picture Dvali and collaborators 
~\cite{Dvali:2010jz,Dvali:2010ns,Dvali:2011nj} make the
even more striking proposal that this feature may be true for a large class 
of non-renormalizable field theories. Instead of the normally assumed UV 
completions such theories could in principle have the same protective 
mechanism discussed above. According to the terminology of 
\cite{Dvali:2010jz}, if this happens, the theory is said to classicalize, and 
the process itself is called classicalization. In analogy with black holes, they 
suggest the existence of similar classical solutions called classicalons, whose 
main characteristic would be, just as in the case of black holes, that their size 
grows with the energy. Again, by going to high energies, instead of probing 
small distances, one would end up going to the classical regime. This remarkable 
idea has been exemplified on two cases: one, the effective theory of Goldstone 
bosons, and the other, the Galileon~\cite{Nicolis:2008in}, which is a scalar 
degree of freedom of the massive graviton, as in say~\cite{Dvali:2000hr}. 
In both these cases one has the explicit form for the classicalons. 

In order to understand quantitatively what is going on, 
an elegant and physical picture was used in~\cite{Dvali:2010ns}. By 
starting with a free spherical wave at infinity, the authors compute the distance 
$r_*$ at which the wave gets appreciably perturbed by the non-renormalizable 
interaction in question. In the case of the 
Goldstone Lagrangian one estimates 

\beq
r_*=L_*\left(L_*E\right)^{1/3}
\eeq

\noi
where $L_*\equiv 1/M_*$ is the small distance cutoff (corresponding to the large 
energy cutoff $M_*$) of the theory. Correspondingly,  \cite{Dvali:2010ns} shows
that the momenta relevant for the scattering are simply cut-off by $1/r_*$, instead
being of order E as one would naively expect.

Now, a less imaginative and more scholastic pursuer of this idea would naturally try 
the usual procedure of path integral quantization around the classicalon, and check
the behaviour of the propagator in order to see what goes on. In this note we do 
precisely that for an illustrative example of Goldstone bosons and confirm the 
results of  \cite{Dvali:2010ns}. We hope that simple minded readers (such as us) 
will find this of some use, and that this appealing program will get some modest
 boost. Besides this, we have nothing further to add to elucidate this picture 
beyond what was done up to now. For a remarkably clear discussion of 
classicalization we refer the reader to~\cite{Dvali:2011nj}.

\section{The solution}

We will study the  d-dimensional Euclidean scalar field theory with 
derivative couplings 

\beq
{\cal L}=\frac{1}{2}\left(\partial\Phi\right)^2+\frac{L_*^d}{4}\left(\partial\Phi\right)^4
\eeq

\noi
This model has solutions of the classical equation of motion with 
a source $Q$

\beq
\partial_\mu\left(\partial_\mu\Phi_{cl}\left(1+L_*^d\left(\partial\Phi_{cl}\right)^2\right)\right)
=Q\delta^d(x-y)
\eeq

The ansatz

\beq
\partial^x_\mu\Phi_{cl}(x)=\frac{\left(x-y\right)_\mu}{\left|x-y\right|}
L_*^{-d/2}f\left(\frac{\left|x-y\right|}{r_0(Q)}\right)
\eeq

\noi
with the definition ($\Omega_d$ is a solid angle in d-dimensions)

\beq
r_0(Q)=\left(\frac{QL_*^{d/2}}{\Omega_d}\right)^{\frac{1}{d-1}}
\eeq

\noi
reduces the equation of motion to

\beq
f(\rho)+f(\rho)^3=1/\rho^{d-1}
\eeq
This is solved simply by ( one should mention that this is the real solution ) 

\beq
f(\rho)=f_+(\rho)-f_-(\rho)
\eeq

\noi
with

\beq
f_\pm(\rho)=\left(\sqrt{\left(\frac{1}{3}\right)^3+
\left(\frac{1}{2\rho^{d-1}}\right)^2}\pm
\frac{1}{2\rho^{d-1}}\right)^{1/3}\\
\eeq
Notice that among all these solutions there is also the trivial ($Q=0$) one.

The physical interpretation of $r_0$ is straightforward: it separates the 
long distance regime when our instanton-like solution
dies off and a short distance behaviour when it grows with r. In this sense, 
it is analogous to $r_*$ of the static classicalon, which as shown 
in~\cite{Dvali:2010ns} is a measure of the cross section ($r_*$ is a distance 
when a free wave at infinity gets perturbed). The change of notation is for 
the sake of emphasizing a different, instanton-like nature of our
solution, needed in order to derive a propagator and an effective interaction. 
In what follows we restrict ourselves to the former; the latter, more involved 
discussion, is left for the future.

\section{The path integral}

The generating functional is

\beq
Z[J]=\int{\cal D}\Phi\exp{\left(-\int d^dx\left({\cal L}-J\Phi\right)\right)}
\eeq

We expand around classical solutions of the equation of motion

\beq
\Phi=\Phi_{cl}+\Phi_q
\eeq

\noi
After repeated use of partial integration and equation of motion, one gets

\beq
\int d^dx{\cal L}(\Phi)\to\int\left({\cal L}(\Phi_{cl})-\Phi_qQ\delta^d(x-y)+
\frac{1}{2}\Phi_q\hat O_{r_0,y}\Phi_q+{\cal L}_{int}(\Phi_q)\right)
\eeq

\noi
with the quadratic operator 

\beq
\label{oq}
\hat O_{r_0,y}=-\partial_\mu\left(\partial_\mu+L_*^d\left(2\partial_\mu\Phi_{cl}\partial_\nu\Phi_{cl}+
\delta_{\mu\nu}(\partial\Phi_{cl})^2\right)\partial_\nu\right)
\eeq
where $r_0$ is the size of the classicalon, and y denotes its position.
Finally we redefine

\beq
\label{jredef}
J(x)\Phi(x)\to(J(x)-Q\delta^d(x-y))\Phi_q(x)
\eeq

\noi
so that the path integral becomes as usual

\bea
Z[J]&\propto&\int dr_0 d^dy f(r_0,y)e^{-S[\Phi_{cl}(r_0,y)]}\left(\det{\hat O_{r_0,y}}\right)^{-1/2}\\
&\times&
\exp{\left(-S_{int}[\delta/\delta J]\right)}\exp{\left(\frac{1}{2}\int d^dx\int d^dzJ(x)
\Delta_{r_0,y}(x,z)J(z)\right)}\nonumber
\eea

\noi
where $r_0$ and $y$ are the collective coordinates, $f(r_0,y)$ is a corresponding measure, 
and $\Delta_{r_0,y}(x)$ is the propagator, i.e. the inverse of the operator (\ref{oq}).

Now it is easy to see how the amplitude looks like. For example, considering for illustration 
only the part of the 4-point Green's function coming from the quartic interaction, we get

\bea
\label{a4}
G_4&=& L_*^d\int dr_0 dy f(r_0,y)e^{-S[\Phi_{cl}(r_0,y)]}\left(\det{\hat O_{r_0,y}}\right)^{-1/2}\\
&\times&\left[\partial_\mu^z\Delta_{r_0,y}(x_1,z)\partial_\mu^z\Delta_{r_0,y}(x_2,z)
\partial_\nu^z\Delta_{r_0,y}(x_3,z)\partial_\nu^z\Delta_{r_0,y}(x_4,z)\right.\nonumber\\
&+&\left.(x_2\leftrightarrow x_3)+(x_2\leftrightarrow x_4)\right]
/\int dr_0 dy f(r_0,y)e^{-S[\Phi_{cl}(r_0,y)]}\left(\det{\hat O_{r_0,y}}\right)^{-1/2}\nonumber
\eea

All one needs to do now is to find out the propagator from

\beq
\hat O_{r_0,y}^x\Delta_{r_0,y}(x,z)=\delta^d(x-z)
\eeq

In order to ease the reader's pain and without the loss of generality we 
choose hereafter $y=0$.

Then the operator (\ref{oq}) can be rewritten as 

\beq
\hat O_{r_0}=-\frac{1}{r^{d-1}}\frac{\partial}{\partial r}
\left(\left[1+3f^2(r/r_0)\right]r^{d-1}
\frac{\partial}{\partial r}
\right)\nonumber+\left[1+f^2(r/r_0)\right]\frac{L^2}{r^2}
\eeq

\noi
with 

\beq
L^2=-\frac{1}{2}\left(x_\mu\partial_\nu-x_\nu\partial_\mu\right)^2
\eeq

For the $d$-dimensional radial coordinate small the operator becomes

\beq
r\to 0:\hskip 1cm\hat O_{r_0}(r)\to-\left(\frac{r_0}{r}\right)^{2(d-1)/3}\left(3\frac{\partial^2}{\partial r^2}+\frac{d-1}{r}\frac{\partial}{\partial r}-\frac{L^2}{r^2}\right)
\eeq

\noi
while for large distances we have the usual Laplace

\beq
r\to\infty:\hskip 1cm \hat O_{r_0}(r)\to-\left(\frac{\partial^2}{\partial r^2}+\frac{d-1}{r}\frac{\partial}{\partial r}-\frac{L^2}{r^2}\right)
\eeq

In a very crude approximation one could say, that the operator becomes infinite for $r\ll r_0$ and Laplacian for 
large $r$. Physically this implies that the momentum space propagator is given roughly by the 
Heaviside function

\beq
\tilde\Delta_{r_0}(p)\approx\frac{1}{p^2}\Theta(1/r_0-p)
\eeq

\noi
so that the maximum available momentum is $p_*=1/r_0$. 
From (\ref{a4}), the amplitude thus becomes

\beq
A_4\to L_*^dp_*^4
\eeq

\noi
as predicted by classicalization. The relevant momenta that characterize high-energy 
scatterings become small, corresponding to the inverse size of the classicalon. 

\section{Summary and outlook}

A great success of the UV completion of the Fermi theory in 
the form of the electro-weak Standard Model, has almost created a dogma 
that this is the only road towards making sense out of non-renormalizable 
theories with dimensionful couplings. The search for the UV completion degrees
of freedom in case of gravity has led eventually to strings, whose impact on 
high energy physics has been enormous. And yet, as argued in~\cite{Dvali:2010bf}, 
it could be that gravity may need no such completion, for it has a built-in 
protective mechanism in the form of classical black holes. Scattering two particles
with energies of say a mammoth, will simply produce black holes with a radius 
proportional the energy of the mammoth, and one will never arrive at
short distances.  In other words, it is not clear a priori that gravity must be 
modified.

Could this be true of other, non gravitational, effective non-renormalizable theories? 
If such theories are energy self-sourced, 
Dvali et al~\cite{Dvali:2010jz,Dvali:2010ns,Dvali:2011nj} argue that the answer
is affirmative. In~\cite{Dvali:2010ns} this is shown by studying the distance at 
which scattering takes place. In this note, we offer a canonical path integral 
discussion of this important result, with a hope that this may help one to further 
grasp this fascinating subject. This is our main apology for deciding to make 
our work public without waiting for a more complete analysis of the issues addressed.

\section*{Acknowledgements}

We are grateful to Gia Dvali for useful discussions and strong encouragement to make 
this work public. We also thank Alejandra Melfo for discussions and careful reading of 
the manuscript. B.B. acknowledges the Slovenian Research Agency for support. A. M. 
would like to thank ICTP for its hospitality and support.

\end{document}